# Characterization of FBK small-pitch 3D diodes after neutron irradiation up to $3.5\times10^{16}$ $n_{eq}$ cm$^{-2}$

**Roberto Mendicino**[a,b], **Maurizio Boscardin**[c,b], **Gian-Franco Dalla Betta**[a,b*]

[a] *Dipartimento di Ingegneria Industriale, Università di Trento,*
  *Via Sommarive 9, 38123 Trento, Italy*
[b] *TIFPA INFN,*
  *Via Sommarive 14, 38123 Trento, Italy*
[c] *Fondazione Bruno Kessler,*
  *Via Sommarive 18, 38123 Trento, Italy*

 *E-mail*: gianfranco.dallabetta@unitn.it

ABSTRACT: We report on the characterization by a position resolved laser system of small-pitch 3D diodes irradiated with neutrons up to an extremely high fluence of $3.5\times10^{16}$ $n_{eq}$ cm$^{-2}$. We show that very high values of signal efficiency are obtained, in good agreement with the geometrical expectation based on the small values of the inter-electrode spacings, and also boosted by charge multiplication effects at high voltage. These results confirm the very high radiation tolerance of small-pitch 3D sensors well beyond the maximum fluences expected at the High Luminosity LHC.

---

[*] Corresponding author.

**Contents**

**1. Introduction**

3D pixels are the most radiation-hard silicon sensors, owing to their structure which allows for a very short distance between the vertical electrodes [1]. This makes them a promising option for the innermost layers of tracking detectors at the High Luminosity LHC (HL-LHC) [2], where, due to the very high luminosity ($5\times10^{34}$ cm$^{-2}$ s$^{-1}$), it is expected to have event pile-up as high as 200 events/bunch crossing. In order to meet the requirements of very high hit-rate capabilities, increased pixel granularity, and extreme radiation hardness, future 3D pixels will have all geometries significantly downscaled as compared to the existing ones, and, in particular, a smaller pitch (e.g., 50×50 or 25×100 μm$^2$), compatible with the new read-out chips designed by the RD53 Collaboration [3], and reduced active thickness (~100 μm).

In view of the HL-LHC "Phase2" detector upgrades, a new generation of 3D pixel sensors has been developed in the past few years by the INFN-FBK collaboration [4]. Devices are fabricated using a single-sided technology on Si-Si Direct Wafer Bonded 6" substrates [5]. They exhibit very good electrical characteristics both before and after irradiation [6]. Beam test results of the first pixel prototypes are encouraging, with hit efficiency values of ~99% (before irradiation) and ~97% (after irradiation to $1\times10^{16}$ n$_{eq}$ cm$^{-2}$) [7].

TCAD simulations incorporating advanced radiation damage models [8] anticipate that charge trapping effects are strongly attenuated in these small-pitch 3D sensors, thus enabling for a good charge collection efficiency (CCE) even after very large irradiation fluences [5]. Moreover, due to the small values of the inter-electrode distance, the electric field intensities can reach values high enough to lead to impact ionization effects and charge multiplication, but also to risk of early breakdown. In order to investigate these effects, and provide direct feedback to simulations, 3D pixel assemblies are not the most suitable vehicle, due to the limited radiation tolerance and relatively coarse charge signal resolution of the read-out chips. In this respect, 3D strips would provide a better solution, since they can be coupled to LHC-compatible read-out electronics (e.g., the ALIBAVA system) without need of bump bonding, and they can be irradiated separately from the read-out chip up to very large fluences [9-12].

As an alternative, we have used 3D diodes, e.g., small (~2 mm$^2$) arrays of basic 3D cells with all columns of the same doping type shorted together, reproducing the same layout details of their parent pixel sensors. While 3D diodes are perfect to investigate the electrical properties, their large capacitance (tens of pF) [5] and leakage current after heavy irradiation (several μA at -10°C) [6] are responsible for high noise in functional measurements (~several ke$^-$ rms). This would make measurements with a β radioactive source quite critical, since the amount of charge released by a minimum ionizing particle in the thin active layer is relatively small (the most probable value in 130 μm is ~9500 ke$^-$). On the contrary, the impact of high noise can be minimized in case of laser tests, by increasing the laser intensity and by averaging the results over a large number of measurements.

The results of position resolved laser tests performed on 3D diodes of different geometries after irradiation with neutrons up to an extremely large fluence of 3.5×10$^{16}$ n$_{eq}$ cm$^{-2}$ are reported in this paper, which is organized as follows. Devices, irradiations and tests are described in section 2. Experimental results are reported and discussed in section 3. Conclusions follow in section 4.

## 2. Devices, irradiation and tests

In this study, we have used 3D diodes belonging to the first batch of small-pitch 3D pixel sensors fabricated at FBK using a single-sided process [5]. The active layer is a high-resistivity, p-type Float Zone silicon wafer of 130 μm thickness (t$_{FZ}$), directly bonded on top of a low-resistivity, p-type handle wafer of 500 μm thickness. The schematic cross-section of the device is shown in Fig. 1(a), with details of one n$^+$ column and one p$^+$ column, which have different depths. The former stops at ~20 μm from the handle wafer to avoid early breakdown, whereas the latter penetrates into the handle wafer, so as to allow for sensor bias from the back side. Both types of columns are partially filled with doped poly-Si and then passivated. The n$^+$ columns are contacted by metal on the front side, and are isolated at the surface by using a p-spray layer [5].

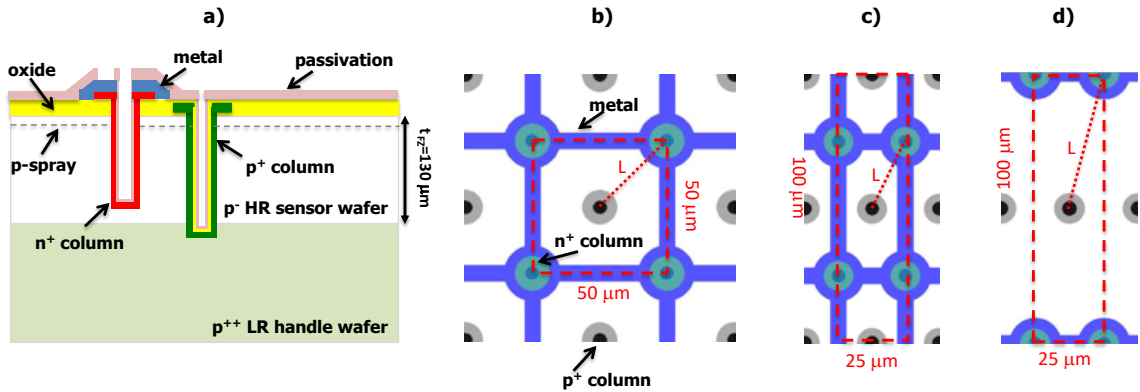

**Figure 1.** a) Schematic cross section of the 3D diodes (not to scale). Layouts of the different geometries under test, corresponding to the regions of interest for position resolved laser measurements: b) 50×50, c) 25×100(2E), and d) 25×100(1E).

Figures 1(b,c,d) show some layout details of the 3D diodes under test, corresponding to the regions of interest for the position resolved laser system. The devices feature three different small-pitch geometries and inter-electrode spacings (L): 50×50 μm$^2$ with one read-out electrode (1E),

with L~35 μm, and 25×100 μm² with either two (2E) or one (1E) read-out electrodes. In the former the basic 3D cell is 25×50 μm² with L~28 μm, in the latter L is much larger (~51 μm).

Sensors were irradiated with neutrons at the TRIGA Mark II reactor at JSI (Ljubljana, Slovenia) to three different fluences: $1\times10^{16}$, $2\times10^{16}$, and $3.5\times10^{16}$ $n_{eq}$ cm⁻². Devices were irradiated without bias (i.e., with floating terminals). All irradiated samples were stored in a freezer to avoid annealing. However, although the measurement setup required to keep them at room temperature for only a few minutes, the sensors might have experienced some annealing during irradiation, when the temperature is not controlled.

For functional tests, we have used a vacuum chamber allowing to operate the samples at low temperature down to -10°C without any problem with condensation. The samples are biased from the back side by placing the silicon dice on a metallic thermal finger covered with a high-conductivity silicon layer that improves the planarity and the electrical contact. A PT100 RTD is also connected to the thermal finger to monitor the temperature in close proximity to the 3D diodes. The signals are read-out from the front side by contacting the probing pads by microneedles. The read-out chain includes a low-noise charge amplifier and a shaper. By using a position resolved pulsed laser (wavelength 1064 nm, nominal pulse width 40 ps), we have collected two-dimensional maps of the relative signal intensity across adjacent 3D basic cells at different bias voltages, from 25 V to 225V in steps of 25 V, or until breakdown occurs. The laser pulses are focused on the sensor surface, where the laser spot has a Gaussian spatial distribution with a standard deviation of about 5 μm. The laser is moved by PC-controlled motorized stages in both X- and Y- directions with ~1 μm precision. Measurements reported in Section 3 were performed using a step size of 3 μm in both X and Y directions, and the maps of the signals were finally interpolated onto 1 μm grids.

The signals acquired on irradiated samples were normalized to the maximum signals acquired on non-irradiated ones, in order to estimate the signal efficiency. The dependence of the light absorption coefficient (α) on the irradiation fluence has been accounted for. Following [13], the value of α on dependence of fluence (ϕ), wavelength (λ), and temperature (T) has been calculated as in Eqn.1:

$$\alpha(\phi, \lambda, T) = \alpha_0(\lambda, T) \cdot \left[1 + \frac{\phi}{\phi_{abs}(\lambda)}\right] \quad (1)$$

where $\alpha_0$ is the absorption coefficient before irradiation, which has been calculated at -10°C according to the model proposed in [14] (but also cross-checked using other models), and the coefficient $\phi_{abs}$ (at λ=1064 nm) has been assumed to be $3.37\pm0.36\times10^{16}$ cm⁻² [13].

The signal normalization is not straightforward, since it is based on the ratio of the collected charge, which in turn depends on the different amounts of charge released by the laser within the relatively thin active layer ($t_{FZ}$=130 μm is much smaller than the light absorption length at 1064 nm, ~1mm) in samples irradiated at different fluences and in the non-irradiated ones. For the calculation of the charge in different samples we used Eqn. 2 based on Lambert-Beer law:

$$Q(\phi, \lambda, T, t_{FZ}) = Q_0 \cdot [1 - exp(-\alpha(\phi, \lambda, T) \cdot t_{FZ})] \quad (2)$$

where $Q_0$ is a proportionality factor (that is eventually canceled in the signal normalization).

Table 1 summarizes all relevant values used in the calculations. The signals measured in irradiated samples were divided by the scaling factor shown in the last column of Table 1, which corresponds to the ratio of the charge released by the laser in irradiated and non-irradiated samples, and finally normalized to the maximum signals measured before irradiation.

| ϕ ($10^{16}$ $n_{eq}/cm^2$) | α @ -10°C ($cm^{-1}$) | Charge in $t_{Fz}$ ($10^{-2}$ $Q_0$) | Scaling factor |
|---|---|---|---|
| 0 | 5.88 | 7.36 | n.a. |
| 1.0 | 7.62(+0.21/-0.17) | 9.43(+0.25/-0.20) | 1.28(+0.04/-0.02) |
| 2.0 | 9.36(+0.42/-0.33) | 11.46(+0.48/-0.39) | 1.56(+0.06/-0.05) |
| 3.5 | 11.98(+0.73/-0.59) | 14.42(+0.81/-0.66) | 1.96(+0.11/-0.09) |

**Table 1.** Summary of data relevant to 1064 nm light absorption at -10°C and charge released in the 130-μm thick active layer before irradiation and after irradiation at different fluences. The uncertainties on the value of $\phi_{abs}$ were propagated to the other quantities, reaching at most ~5% in the scaling factor.

## 3. Results and discussion

### 3.1 3D diode of 50×50 type

Figure 3 shows the two-dimensional maps of the signal efficiency (SE) measured at three different voltages in the sample irradiated at $1\times10^{16}$ $n_{eq}$ $cm^{-2}$. At 25 V the SE is low everywhere, then it gradually increases reaching high values at 125 V, and eventually exceeding 100% at 225 V. The SE is maximum in the regions close to the $p^+$ columns, whereas it is minimum in the regions covered by metal. In the latter, signals are observed anyway due to the non-negligible size of the light spot size (the metal grid is only 5 μm wide) and to the fact that columns are partially empty, so that metal does not perfectly cover/shield their center (see details of the structure in Fig. 1a). However, it is not possible to estimate the amount of charge released in these regions, so the efficiency values therein are not significant. The maps of signal efficiency and their evolution with voltage are similar also in samples irradiated at larger fluences; however, larger voltages are required before the efficiency reaches high values, as expected from the increased radiation damage effects both in terms of larger depletion voltage and stronger charge trapping.

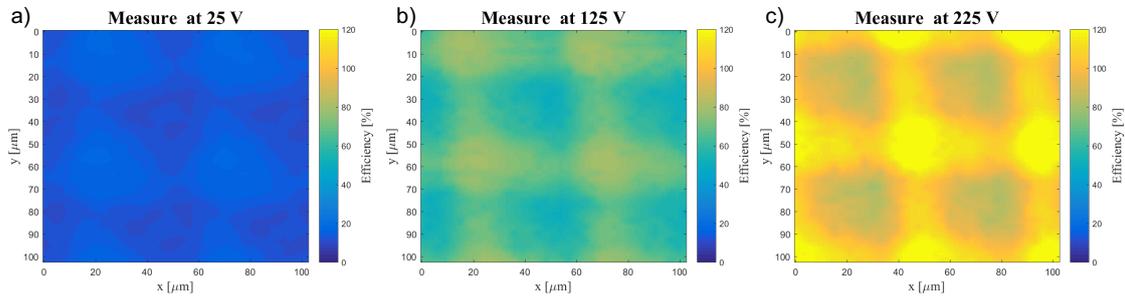

**Figure 3.** Two-dimensional maps of signal efficiency in a 3D diode of 50×50 type irradiated at $1\times10^{16}$ $n_{eq}$ $cm^{-2}$ at three bias voltages: a) 25 V, b) 125 V, and c) 225 V.

The behavior of all devices at different voltages can be better appreciated from Fig. 4: slices a, b, and c show the signal efficiency as a function of distance along a line connecting the center of a $n^+$ column (0) to the center of a $p^+$ column (~35 μm) at different bias voltages and for the three considered fluences. The increase of the SE with voltage is quite uniform at all positions, but more pronounced closer to the $p^+$ column, especially at the largest voltages. The average values of different curves from Fig. 4(a,b,c), excluding those relevant to the region covered by metal, are plotted in Fig. 4d as a function of voltage, with error bars representing the standard deviations. It can be seen that SE monotonically increases with voltage at all fluences, eventually reaching remarkably high values that exceed 100% at $1\times10^{16}$ $n_{eq}$ cm$^{-2}$ and $2\times10^{16}$ $n_{eq}$ cm$^{-2}$, and 80% at $3.5\times10^{16}$ $n_{eq}$ cm$^{-2}$.

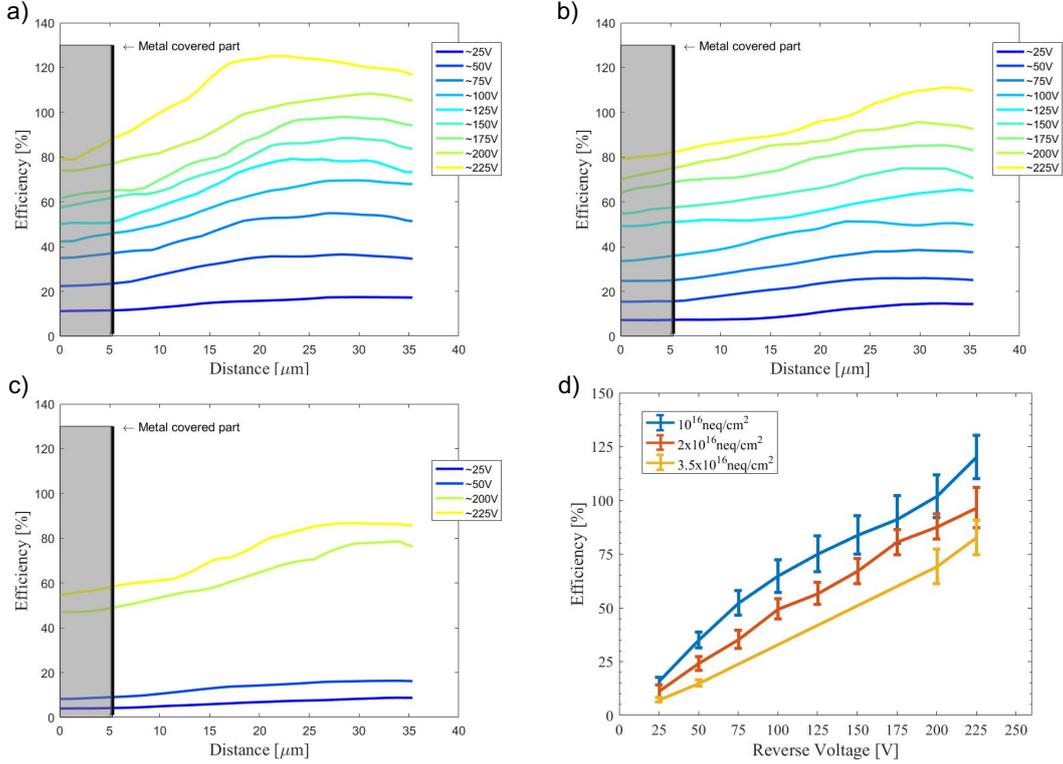

**Figure 4.** Signal efficiency as a function of distance along a line connecting the center of a n$^+$ column (0) to the center of a p$^+$ column (~35 μm) at different bias voltages in a 3D diode of 50×50 type irradiated at different fluences: a) $1\times10^{16}$ $n_{eq}$ cm$^{-2}$, b) $2\times10^{16}$ $n_{eq}$ cm$^{-2}$, and c) $3.5\times10^{16}$ $n_{eq}$ cm$^{-2}$ ; d) corresponding signal efficiency as a function of voltage at the three fluences (data represent the average of the values of slices a, b, c, excluding the regions covered by metal, whereas the error bars are the standard deviations).

### 3.2 3D diode of 25×100 (2E) type

Fig. 5 shows the two-dimensional maps of signal efficiency measured at the three fluences at a reverse bias of 150 V. This comparison clearly illustrates the increasing impact of radiation damage on the efficiency as the fluence is increased.

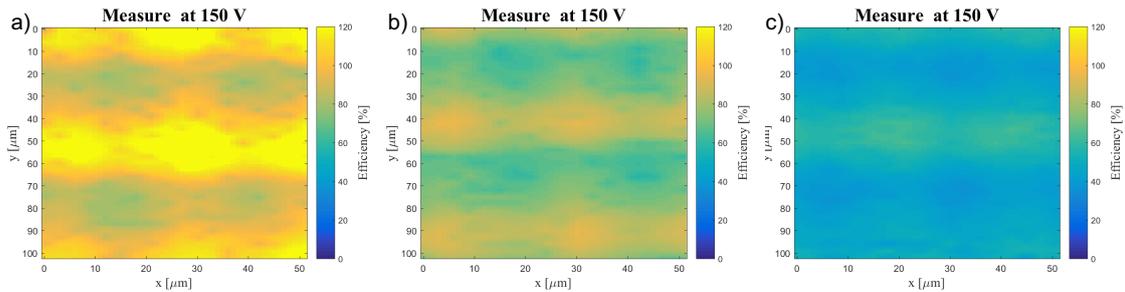

**Figure 5.** Two-dimensional maps of signal efficiency at 150 V bias voltage in a 3D diode of 25×100(2E) type irradiated at: a) $1\times10^{16}$ $n_{eq}$ cm$^{-2}$, b) $2\times10^{16}$ $n_{eq}$ cm$^{-2}$, and c) $3.5\times10^{16}$ $n_{eq}$ cm$^{-2}$.

– 5 –

A more comprehensive description of the charge collection behavior can be observed in Fig. 6, where slices a, b, and c show the signal efficiency as a function of distance along a line connecting the center of a n$^+$ column (0) to the center of a p$^+$ column (~28 μm) at different bias voltages and for the three considered fluences. The results are qualitatively similar to those shown for the previous geometry. As for the average efficiencies of Fig. 6d, the shapes of the curves are slightly different from those of the 50×50 device, in that they exhibit a quasi-saturation trend, likely ascribed to the smaller value of L which allows for a smaller depletion voltage. Note that in the sample irradiated at 1×10$^{16}$ n$_{eq}$ cm$^{-2}$, the knee in the curve at about 50 V is followed by a smooth rise, which is compatible to the early breakdown occurring in this device at about 150 V.

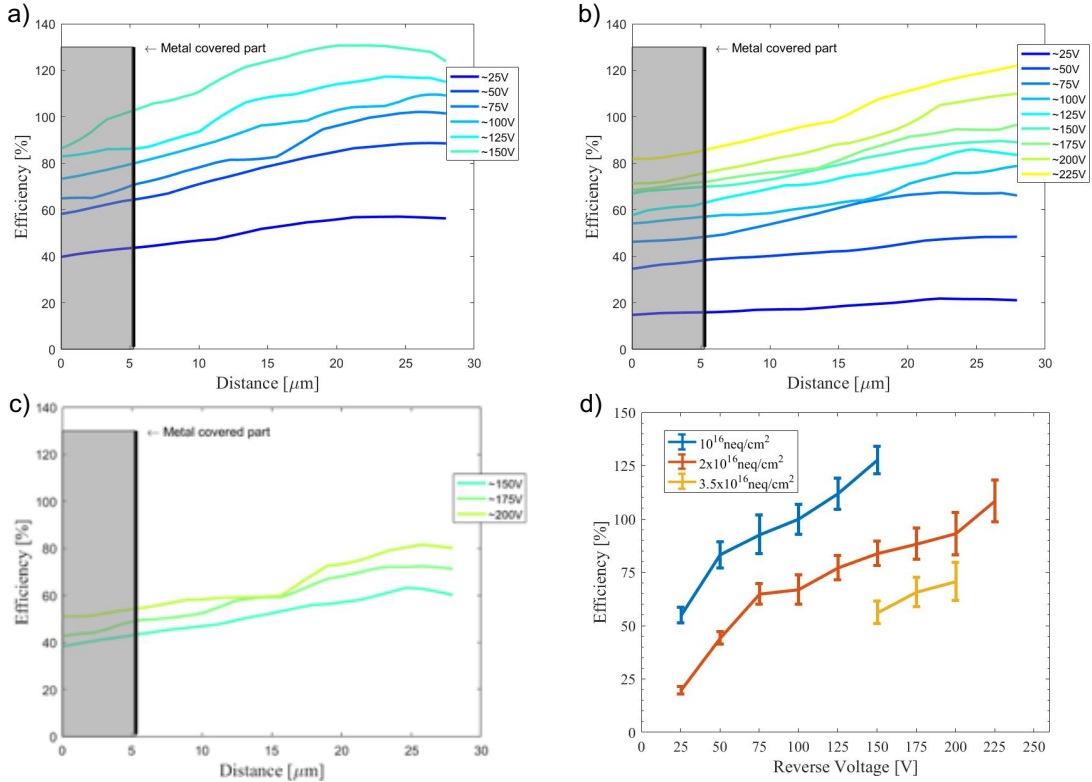

**Figure 6.** Signal efficiency as a function of distance along a line connecting the center of a n$^+$ column (0) to the center of a p$^+$ column (~28 μm) at different bias voltages in a 3D diode of 25×100(2E) type irradiated at different fluences: a) 1×10$^{16}$ n$_{eq}$ cm$^{-2}$, b) 2×10$^{16}$ n$_{eq}$ cm$^{-2}$, and c) 3.5×10$^{16}$ n$_{eq}$ cm$^{-2}$ ; d) corresponding signal efficiency as a function of voltage at the three fluences (data represent the average of the values of slices a, b, c, excluding the regions covered by metal, whereas the error bars are the standard deviations).

### 3.3 3D diode of 25×100 (1E) type

Unfortunately, the sensors irradiated at 1×10$^{16}$ n$_{eq}$ cm$^{-2}$ and 3.5×10$^{16}$ n$_{eq}$ cm$^{-2}$ did not work properly, with early breakdown, so it was possible to perform the measurements only on the one irradiated at 2×10$^{16}$ n$_{eq}$ cm$^{-2}$. Relevant results are summarized in Fig.7a, which shows the signal efficiency as a function of distance along a line connecting the center of a n$^+$ column (0) to the center of a p$^+$ column (~51 μm) at different bias voltages, until breakdown. Here the efficiencies are sizably smaller than in the other devices, due to the smaller value of L, and the region showing the highest values is that closer to the n$^+$ column, with a secondary peak close to the p$^+$ column.

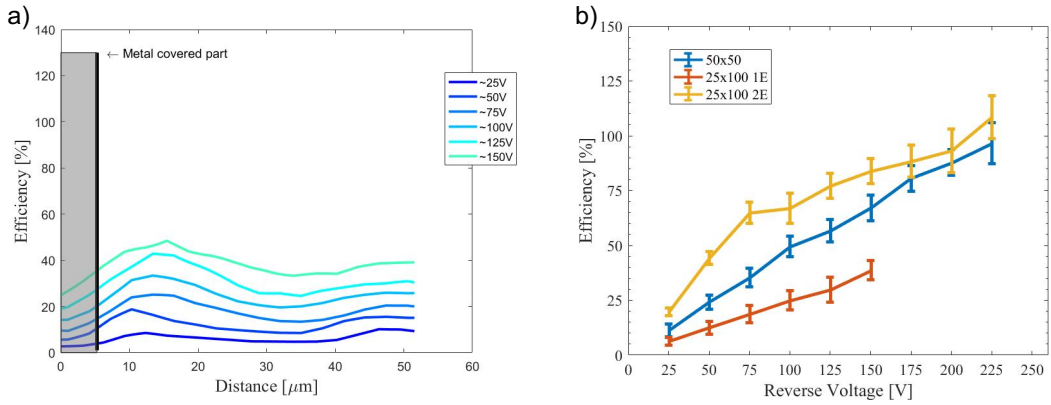

**Figure 7.** a) Signal efficiency as a function of distance along a line connecting the center of a $n^+$ column (0) to the center of a $p^+$ column (~51 μm) at different bias voltages in a 3D diode of 25×100(1E) type irradiated at $2\times10^{16}$ $n_{eq}$ cm$^{-2}$; b) signal efficiencies as a function of voltage in 3D diodes of different geometries irradiated at $2\times10^{16}$ $n_{eq}$ cm$^{-2}$.

## 4. Discussion and conclusion

The results are in quite a good agreement with TCAD simulation predictions [5, 15]. Taking as a benchmark the $2\times10^{16}$ $n_{eq}$ cm$^{-2}$ fluence, for which data from all geometries are available, the signal efficiencies as a function of voltage are compared in Fig. 7b. Up to a voltage of 150 V, SE's properly scale with the inter-electrode spacing, L [1]. The maximum SE is obtained in the 25×100(2E) device, ~85% at 150 V, with a quasi-saturation trend observed already at 75 V. At 150 V, in the 50×50 device, the SE reaches a remarkable value of ~67%, whereas it is smaller but still pretty good (~39%) in the 25×100(1E) sample. In order to limit the power dissipation in the final application, operating at this voltage still seems to offer a sufficiently good trade-off with signal efficiency. At larger voltages, impact ionization effects lead to charge multiplication, boosting the signal efficiency to 100% and beyond in both the 25×100(2E) and 50×50 devices. This is not evident in the 25×100(1E) sample where electrical breakdown prevents from reaching the much larger voltages that would likely be necessary for charge multiplication with L~51 μm.

The results are also found to be comparable to the experimental data reported by other groups for small-pitch 3D strip sensors characterized with a β source or in beam tests. As an example, with reference to the 50×50 geometry, 90%(80%) SE at 175 V after $1\times10^{16}$ ($1.5\times10^{16}$) $n_{eq}$ cm$^{-2}$ is reported in [11]; moreover, 70-75% (depending on the setup) SE at 175 V after $1.7\times10^{16}$ $n_{eq}$ cm$^{-2}$ is reported in [12]. In our samples, at 175 V, the signal efficiency is 90%(80%) after $1\times10^{16}$ ($2\times10^{16}$) $n_{eq}$ cm$^{-2}$, very close to the previous values.

In spite of the uncertainties related to the laser setup, results from the present study confirm the very high radiation tolerance of small-pitch 3D sensors up to extremely high fluences, and are encouraging in view of their application to the innermost tracking layers at the High Luminosity LHC. Of course, it should be proved that a high signal efficiency corresponds to a high hit efficiency in a beam test. In this respect, very promising results were recently reported from CNM samples of 230-μm thickness [16], and should be confirmed on devices of smaller thickness. Further studies are also under way to explain in detail the evolution of the SE with voltage at different positions within the 3D cell, providing useful feedback for the refinement of the bulk radiation damage model.

## Acknowledgments

This project has received funding from the European Union's Horizon 2020 Research and Innovation programme under Grant Agreement no. 654168. This work was also supported by the Autonomous Province of Trento and by the Italian National Institute for Nuclear Physics (INFN).